\documentclass[12pt,preprint,usenatbib]{aastex}

\shorttitle{Radial distributions of the two main-sequence components
  in NGC 1856}

\shortauthors{Li et al.}

\begin{document}
\title{The radial distributions of the two main-sequence components in
  the young massive star cluster NGC 1856} 
\author{Chengyuan Li$^{1}$, Richard de Grijs$^{2,3}$, Licai
  Deng$^{4}$, \and Antonino P. Milone$^{5}$}


\affil{$^{1}$Department of Physics and Astronomy, Macquarie
  University, Sydney, NSW 2109, Australia}
\email{chengyuan.li@mq.edu.au}
\affil{$^{2}$Kavli Institute for Astronomy \& Astrophysics and
  Department of Astronomy, Peking University, Yi He Yuan Lu 5, Hai
  Dian District, Beijing 100871, China}
\affil{$^{3}$International Space Science Institute--Beijing, 1
  Nanertiao, Zhongguancun, Hai Dian District, Beijing 100190, China}
\affil{$^{4}$Key Laboratory for Optical Astronomy, National
  Astronomical Observatories, Chinese Academy of Sciences, 20A Datun
  Road, Chaoyang District, Beijing 100012, China}
\affil{$^{5}$Research School of Astronomy \& Astrophysics, Australian
  National University, Mt Stromlo Observatory, Cotter Rd, Weston, ACT
  2611, Australia}

\begin{abstract}
The recent discovery of double main sequences in the young, massive
star cluster NGC 1856 has caught significant attention. The
observations can be explained by invoking two stellar generations with
different ages and metallicities or by a single generation of stars
composed of two populations characterized by different rotation
rates. We analyzed the number ratios of stars belonging to both
main-sequence components in NGC 1856 as a function of radius. We found
that their number ratios remain approximately unchanged from the
cluster's central region to its periphery, indicating that both
components are homogeneously distributed in space. Through a
comparison of the loci of the best-fitting isochrones with the ridge
lines of both stellar components, we found that both multiple stellar
populations and rapid stellar rotation can potentially explain the
observed main-sequence bifurcation in NGC 1856. However, if NGC1856
were a young representative of the old globular clusters, then the
multiple stellar populations model would not be able to explain the
observed homogeneity in the spatial distributions of these two
components, since all relevant scenarios would predict that the second
stellar generation should be formed in a more compact configuration
than that of the first stellar generation, while NGC 1856 is too young
for both stellar generations to have been fully mixed dynamically. We
speculate that the rapid stellar rotation scenario would be the
favored explanation of the observed multiple stellar sequences in NGC
1856.
\end{abstract}


\section{Introduction}

Current consensus on the formation of star clusters suggests that a
cluster's initial star-formation process approximately resembles a
single burst \citep{Long14a}. However, the discovery of ubiquitous
multiple stellar populations in globular clusters (GCs)
\citep[e.g.,][]{Carr09a,Milo12a,Piot15a}, as well as the extended
main-sequence turn-offs (eMSTOs) in most intermediate-age star
clusters \citep[e.g.,][]{Milo09a,Gira13a} has strongly challenged this
notion. The nature of the eMSTOs in intermediate-age star clusters is
still being debated \citep{Li14a,Bast15a,Goud15a,Bast16a}. The
multiple stellar populations in GCs exhibit significantly different
chemical abundances \citep[e.g.,][]{Kraf79a,Carr09a}, which is in
strong conflict with the expectations from the formation scenario of
coeval stellar populations. For example, \cite{Milo12a} found two
distinct main sequences (MSs) in the GC 47 Tucanae (47 Tuc), which 
can be explained by CN-weak/O-rich/Na-poor/He-poor and 
CN-strong/O-poor/Na-rich-He-rich populations. 
The latter has been speculated to represent a second
stellar generation, which may have originated from material whose
metallicity had been enhanced by the waste products of the
first-generation stars. {\color{black} In addition, \cite{diCr10a}
  studied the morphology of MS, horizontal-branch (HB) and subgiant branch
  (SGB) stars in 47 Tuc using population synthesis. They concluded
  that 47 Tuc may also possess a spread in helium abundance in its
  more evolved stars \citep[for confirmation, see][]{Nata11a}.}
Similar results have been also found for the GC NGC 2808, which shows
triple MSs that can be explained by three stellar populations with
different helium abundances \citep{Piot07a}. Some GCs, including NGC
1851 and NGC 6656, exhibit split subgiant branches (SGBs) in optical
filters which correspond to stellar populations containing different
amounts of heavy elements \citep[see][and references in their Table
  10]{Mari15a}. In addition, the red-giant branches (RGBs) in most GCs
are split into multiple components, especially when ultraviolet (UV)
observations are used \citep[e.g.,][]{Piot15a}.

Based on accurate photometry, marked differences in the dynamics of
different stellar populations have also been found in many
GCs. \cite{Milo12a} found that the second stellar generation in 47 Tuc
(the CN-strong/O-poor/Na-rich population) has a higher central
concentration than the first stellar generation, a result that was
confirmed by \cite{Li14b}. \cite{Mass16a} found two well-populated
RGBs in the GC M3 (NGC 5272), which can be explained by an
N-poor/Na-poor first stellar generation and an N-rich/Na-rich second
stellar generation. The second-generation stars are significantly more
centrally concentrated than the cluster's first-generation
stars. \cite{Lard11a} analyzed seven Galactic GCs based on Sloan
Digital Sky Survey (SDSS) photometric data. They found that all
UV--red RGBs in these GCs are more centrally concentrated than their
UV--blue counterparts, where the UV--red RGB stars are mainly composed
of Na-rich stars, i.e., those stars that should belong to the second
stellar generation.

Numerous scenarios have been proposed to explain the multiple stellar
populations in GCs. Most of these draw on self-pollution of the
intra-cluster gas \citep{Li16a}. The polluters that may explain the
enriched secondary stellar generations are diverse, including rapidly
rotating massive stars \citep{Decr07a}, massive binaries
\citep{deMi09a}, and evolved post-giant-branch stars
\citep{Vent09a}. {\color{black} However, most of these scenarios are
  still associated with sometimes severe problems
  \citep[e.g.,][]{Renz15a,Bast15b}.} All of these scenarios predict
that the second stellar generation should formed in a more
concentrated configuration compared with a cluster's first stellar
generation \citep[]{Renz08a,Vesp13a,Hnau15a,Khal15a}.

The first evidence of possible multiple stellar populations in young
clusters was discovered in the Large Magellanic Cloud (LMC) cluster
NGC 1856 \citep{Corr15a,Milo15a}. It shows an apparent eMSTO, and its
MS splits into two distinct components, which cannot be explained by a
coeval, chemically homogeneous stellar population. Similar discoveries
for other young, massive clusters (YMCs) followed: \cite{Bast16a}
found that the MSTO region of the YMC NGC 1850 is significantly more
extended than predicted for simple stellar populations. Similarly to
NGC 1856, it was found that the YMC NGC 1755 also exhibits an apparent
bifurcation in its MS \citep{Milo16a}.

Although there is growing evidence of the reality of multiple stellar
populations in YMCs, the dynamics of these different stellar
populations are still poorly understood. In this paper, we analyze the
radial profile of the number ratio of the two MS components in NGC
1856. We find that both stellar sequences are not very different in
terms of their spatial distributions. The number ratio of the two
components as a function of radius remains constant. Two equally
likely scenarios could potentially explain the observed multiple MS
components in NGC 1856: the observed multiple sequences can be
explained by either invoking multiple stellar populations of different
ages and metallicities, or a coeval stellar population characterized
by member stars that exhibit two different rotation rates
\citep{Dant15a}. However, if the former explanation were correct, and
if NGC 1856 is indeed the infant stage of an old GC, the
second-generation stars should be initially more centrally
concentrated than their first-generation counterparts
\citep{Renz08a,Vesp13a,Hnau15a,Khal15a}. Since NGC 1856 is very young
\citep[$\sim$300 Myr;][]{Milo15a}, this difference in their radial
distributions should be still apparent, but this prediction is in
conflict with our observations. Our result may indicate that the
observed multiple stellar sequences in NGC 1856 have actually been
produced by a coeval stellar population composed of two components of
stars with different rotation rates. However, we emphasize that it is
still unclear whether the stellar components characterized by
different rotation rates would be different in their spatial
distributions; follow-up research to explore this aspect is required.

This article is arranged as follows. Section \ref{S2} describes our
data reduction. Section \ref{S3} contains the main results of our
analysis. In Section \ref{S4} we discuss the possible underlying
physics that may produce the observed results. Section \ref{S5}
includes our conclusions.

\section{Data Reduction}\label{S2}

We used the photometric catalogs of \cite{Milo15a}, based on
observations with the {\sl Hubble Space Telescope} ({\sl HST}) using
the Ultraviolet and Visual channels of the Wide Field Camera 3
(UVIS/WFC3). The data sets used here, which were observed through the
F336W and F438W filters, were obtained as part of the observational
programs GO-13011 (PI: T. H. Puzia) and GO-13379 (PI:
A. P. Milone).\footnote{The complete data sets include five bands,
  i.e., F336W, F438W, F555W, F656N, and F814W: see Fig. \ref{F1}. In
  this paper, we only use the observations in the F336W and F438W
  bands, because the MS bifurcation in NGC 1856 is most apparent in
  (F336W, F336W--F438W) color--magnitude space.} Relevant information
pertaining to our data set is summarized in Table 1 of
\cite{Milo15a}. Figure \ref{F1} shows the footprints of the
observational fields. The resulting stellar catalog is based on
accurate point-spread-function photometry using a software program
developed by \cite{Ande08a}. Both the zero points and the differential
reddening are well calibrated \cite[for more details, see][]{Milo15a}.

\begin{figure}
\centering
\includegraphics[width=0.8\textwidth]{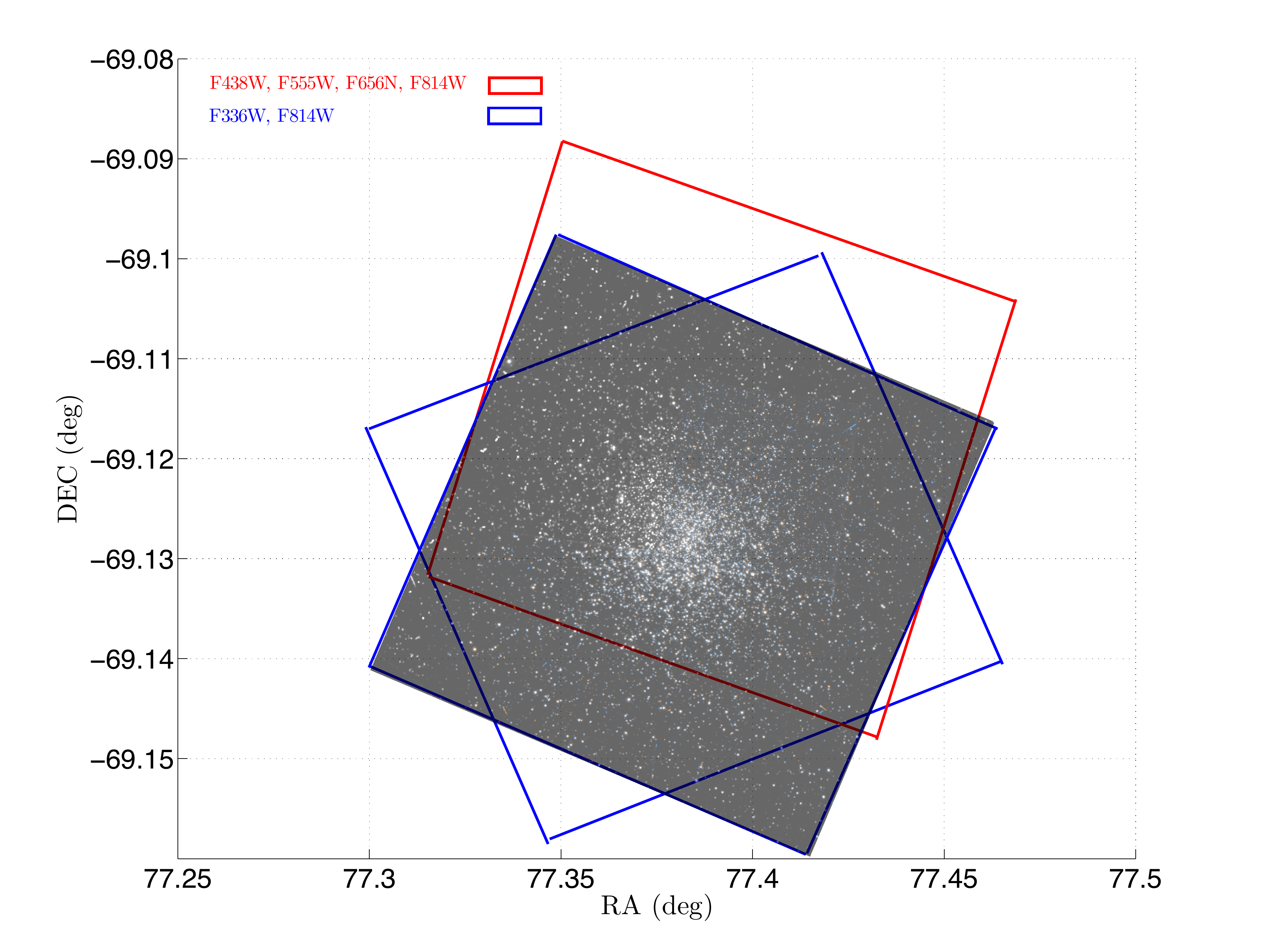}
\caption{Footprints of the NGC 1856 fields. The fields for the F438W,
  F555W, F656N, and F814W observations are based on {\sl HST} program
  GO-13011 (red box); the images in the F336W and F814W filters (blue
  boxes) are from {\sl HST} program GO-13379. For more details, see
  Table 1 of \cite{Milo15a}. The true-color
  image (http://www.wikiwand.com/en/NGC\_1856) clearly shows
  the outlines of the observational fields with respect to the
  position of the cluster.}\vspace{5mm}
\end{figure}\label{F1}

The first step involves the determination of the cluster center
coordinates. Because we will focus on the number ratio of the two
main-sequence components, using the location where the stellar number
density reaches its maximum as cluster center is appropriate. The
method we used to determine the stellar number density center in NGC
1856 is identical to that of \cite{Li16b}: we first divided all stars
into 20 bins along both the right ascension ($\alpha_{\rm J2000}$) and
declination ($\delta_{\rm J2000}$) axes. We then used a Gaussion
function to fit their number density distribution along both axes. The
resulting peaks of the best-fitting Gaussian function in each
direction thus represent the coordinates of the cluster center,
($\alpha_{\rm J2000}=05^{\rm h}09^{\rm m}31.34^{\rm s}$, $\delta_{\rm
  J2000}=-69^{\circ}07'42.07''$). Our result is consistent with the
center coordinates of \cite{Werc11a}.

We next need to define an appropriate cluster size. We calculated the
angular distances to the adopted cluster center for all stars in the
catalog. We divided the observed field into 20 concentric rings, with
radii spanning from $10''$ to $105''$. We then calculated the average
stellar number densities for these 20 rings. This procedure provided
us with a monotonically decreasing number density profile of the stars
in NGC 1856. We simply define the radius where the stellar number
density decreases to about half the central density (the density
within $R=10''$) as the `core' size, $R_{\rm c} = 22.5''$. We used the
region at radii in excess of four times the core size ($R\geq90''$) to
evaluate the corresponding average number density. Since this region
is located far from the cluster center, we treat its average number
density as the `field' level. The radius where the cluster's stellar
number density disappears into the field's stellar population is
defined as the cluster size, $R \approx 70''$ or about three core
radii. We subsequently selected the region with $R\geq75''$ as our
reference field, which will be used for background correction. Using a
Monte Carlo-based approach, we determined that the area of the cluster
region is about 2.3 times larger than that of the referenced field. In
Fig. \ref{F2}, we present the raw (F336W, F336W--F438W)
color--magnitude diagram (CMD) of the cluster region (left panels) as
well as that of the reference field (right panels). An apparent split
in the cluster's MS is found for $19 \leq$ F336W $\leq 21$ mag, which
is shown in the bottom left-hand panel of Fig. \ref{F2}.

\begin{figure*} 
\centering
\includegraphics[width=0.8\textwidth]{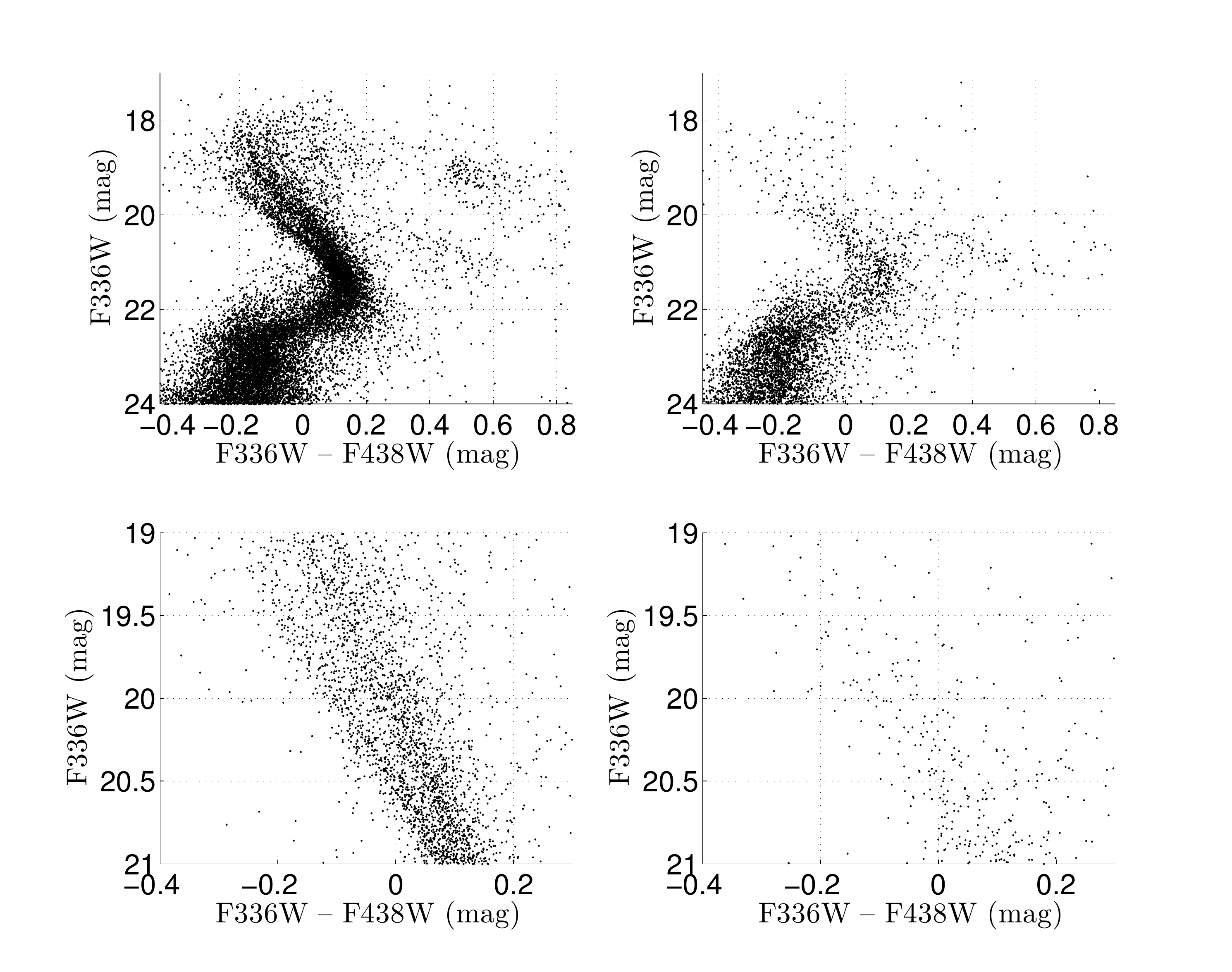}
\caption{(top left) NGC 1856 CMD. (top right) CMD of the reference
  field. (bottom) As the top panels, but highlighting the magnitude
  range $19\le$ F336W $\le21$ mag.}\vspace{5mm}
\end{figure*}\label{F2}

Finally, we calculated the ridge lines describing both MS
components. For both components, we connected the maximum
number-density points across the range $19 \leq$ F336W $\leq 21$ mag.
The resulting ridge lines are shown in Fig. \ref{F3}. We then explored
the dispersions associated with these ridge lines. We used an
exponential function to fit the relationship between the stellar
magnitudes and their corresponding photometric uncertainties, similar
to that used by \citet[][their Fig. 4]{Li13a}. Based on this empirical
function, we defined the boundaries of the dispersion distributions
corresponding to the 3$\sigma$ (standard deviations) levels for both
MS components. Finally, all stars with (a) color--magnitude loci
between the blue and red boundaries (i.e., the blue and red dashed
lines in Fig. \ref{F3}) and (b) F336W magnitudes between 19.25 mag and
20.75 mag were selected as sample stars.

\begin{figure*}
\centering
\includegraphics[width=0.6\textwidth]{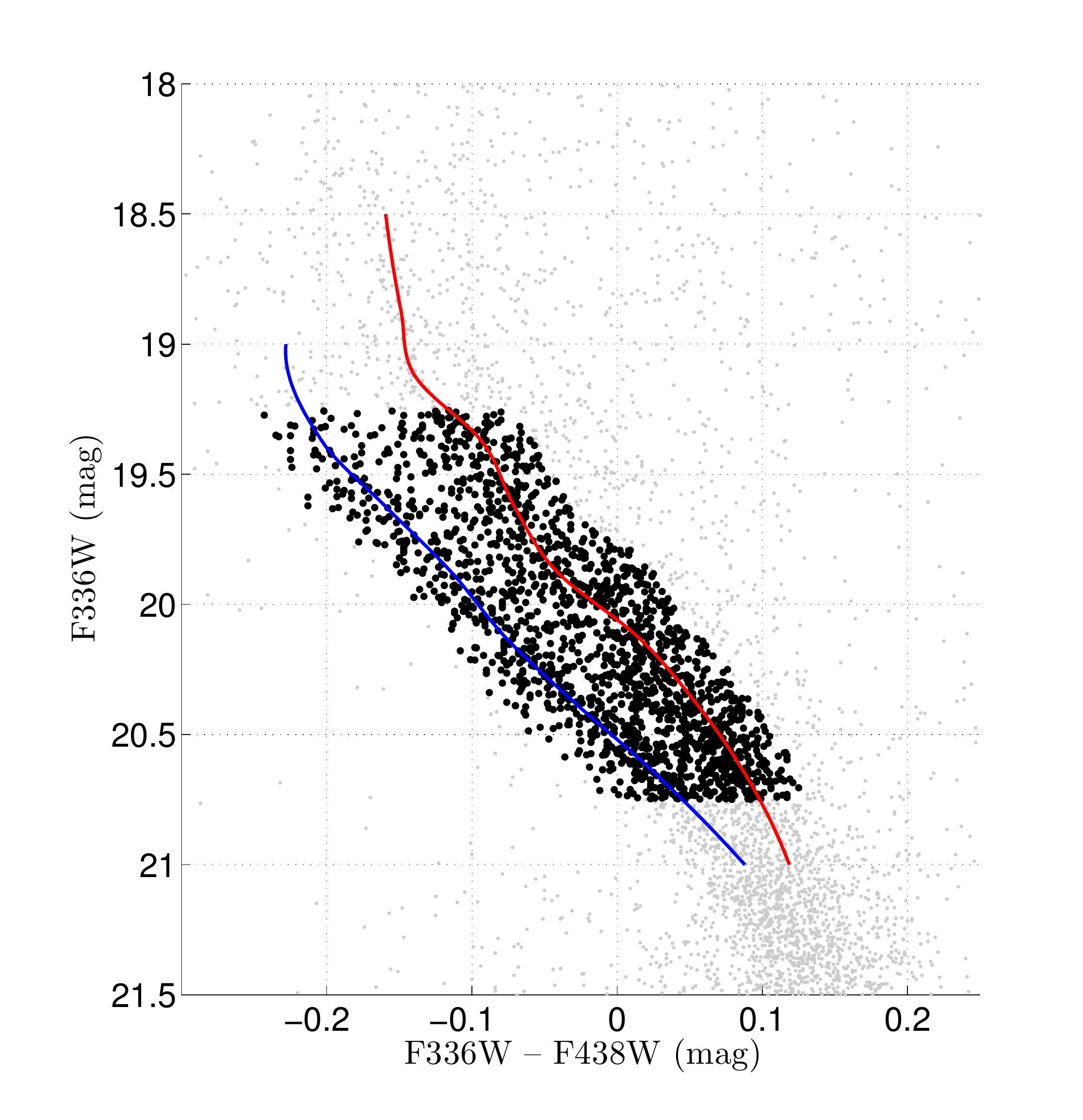}
\caption{CMD of the region containing our sample stars: stars with
  magnitudes in the range $19.25\leq$ F336W $\leq 20.75$ mag and
  colors between the 3$\sigma$ boundaries of the blue and red ridge
  lines are part of our final sample (black dots).}\vspace{5mm}
\end{figure*}\label{F3}

\section{Main Results}\label{S3}

Once we had defined the sample of MS stars, we evaluated the relative
number ratio of stars belonging to both MS components. We used a
similar method to that used in Fig. 1 of \cite{Milo15b}: we projected
both MS components along the vertical axis by defining a parameter
$\Delta^N_{\rm F336W, F438W}$ = $[(C-C_{\rm A})/(C_{\rm B}-C_{\rm
    A})]$ for each star, where $C=F336W - F4388W$ is the star's color
and $C_{\rm A},C_{\rm B}$ are the color residuals corresponding to
ridge lines A and B, respectively. The same approach is also applied
to stars belonging to the reference field. Contamination by field
stars is subsequently reduced based on the $\Delta^N_{\rm F336W,
  F438W}$ distribution of the sample stars. Our resulting number
distribution, $\Delta^N_{\rm F336W, F438W}$, shows two peaks. The
distribution is well-described by two-component Gaussian functions,
\begin{equation}
N(\delta)=N_{\rm A}(\delta)+N_{\rm
  B}(\delta)=a_1e^{-\left(\frac{\delta-b_1}{c_1}\right)^2}+a_2e^{-\left(\frac{\delta-b_2}{c_2}\right)^2},
\end{equation}
where $N(\delta)$ is the number corresponding to a given value of
$\Delta^N_{\rm F336W, F438W}$, and $N_{\rm A}$ and $N_{\rm B}$ are the
contributions from the blue (`A' in Fig. \ref{F4}) and red (`B' in
Fig. \ref{F4}) component stars. The relative number ratio of these two
components can be represented by the ratio of the areas of their
corresponding Gaussian functions,
\begin{equation}
f = \frac{N_{\rm A}}{N_{\rm
    B}}=\frac{\int_{-\infty}^{\infty}a_1e^{-\left(\frac{\delta-b_1}{c_1}\right)^2}{\rm
    d}{\delta}}{\int_{-\infty}^{\infty}a_2e^{-\left(\frac{\delta-b_2}{c_2}\right)^2}{\rm
    d}{\delta}}=\frac{\sqrt{\pi}{a_1}{c_1}/2}{\sqrt{\pi}{a_2}{c_2}/2}
= \frac{a_1c_1}{a_2c_2}.
\end{equation}

The resulting number ratio of stars associated with both components is
$N_{\rm A}$/$N_{\rm B}$=34\%/66\%: see Fig. \ref{F4}. The component-B
stars dominate the stellar numbers \citep[see
  also][33\%/67\%]{Milo15a}.

\begin{figure*}
\centering
\includegraphics[width=0.8\textwidth]{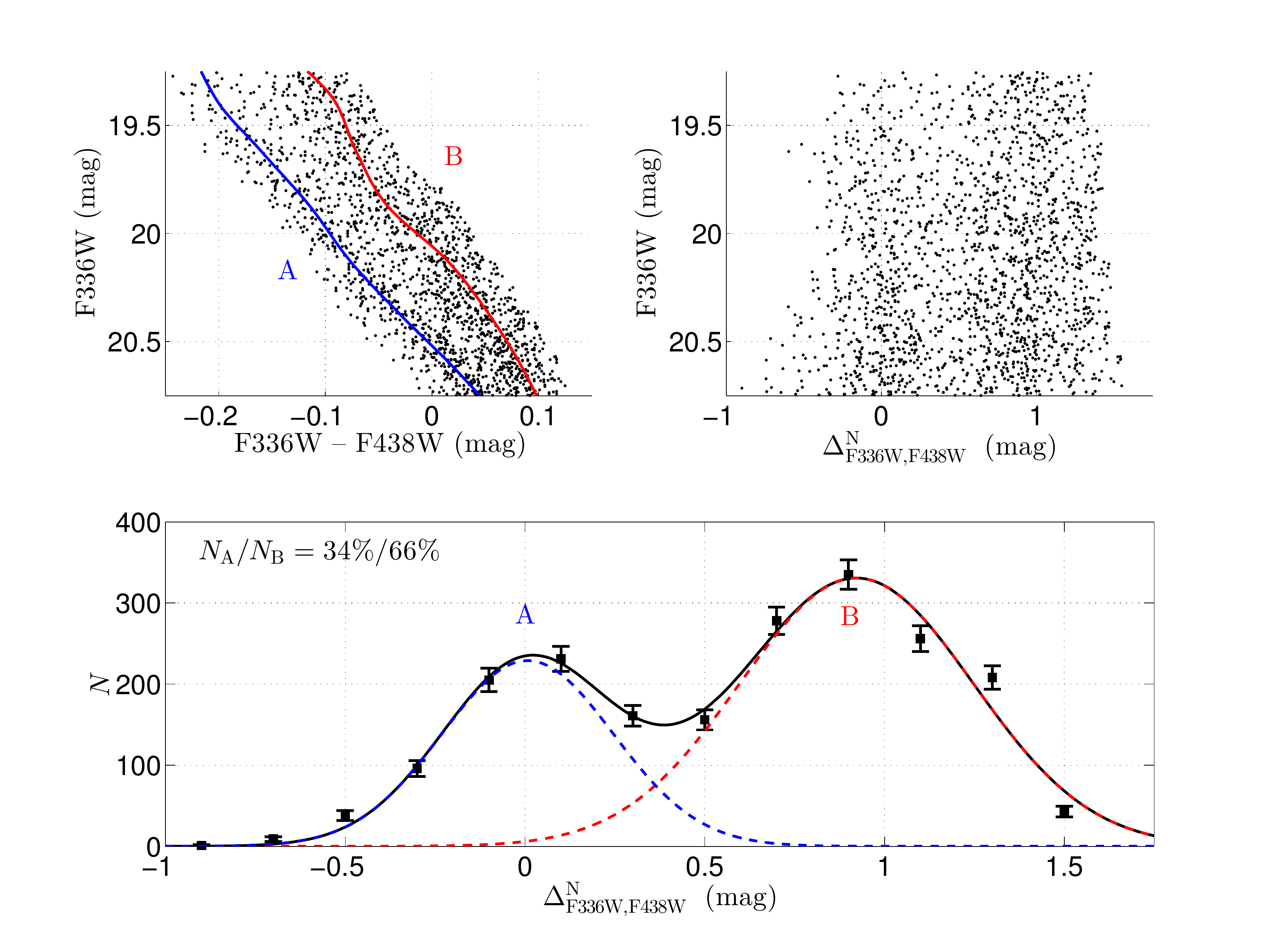}
\caption{Illustration of our calculation of the number ratio of stars
  belonging to both MS components. The top left-hand panel shows the
  CMD of the two MS components and the adopted ridge lines (thick
  lines). The top right-hand panel shows the corresponding
  distributions of the stellar color deviations from these ridge
  lines. The best-fitting double Gaussian profile is shown in the
  bottom panel.}\vspace{5mm}
\end{figure*}\label{F4}

We next examined whether the two stellar components exhibited
different spatial distributions. We first examined the cumulative
profiles of the number distributions of both stellar components. We
divided the sample into seven concentric circles with radii spanning
from $10''$ to $70''$. For each annulus, we performed the same
approach with the stars located at those radii. The cumulative number
distributions of components A and B stars are presented in Fig.
\ref{F5} (top panel). We found that in the 95\% confidence
range,\footnote{In this paper, we simply adopt Poissonian dispersions
  for the numbers of stars in the different components.}, the profiles
of the cumulative number distributions for component-A and -B stars
are not significantly different. This is also reflected by the profile
of their number ratio, $N_{\rm A}/N_{\rm B}$. The value of $N_{\rm
  A}/N_{\rm B}$ for the entire cluster region ($10''\leq{R}\leq70''$)
ranges from 43\% to 51\%, and the uncertainties corresponding to 
the 95\% confidence interval range from $\sim$4\% to 9\%.

We also studied the annular $N_{\rm A}/N_{\rm B}$ profile. We divided
our sample into seven rings with boundaries at 10, 15, 20, 27, 35, and
45 arcsec. These boundaries were adopted to ensure that the stellar
numbers in each ring were similar. As expected, the resulting $N_{\rm
  A}/N_{\rm B}$ profile is highly dispersed compared with the
cumulative profile (see the grey dashed line in the bottom panel of
Fig.  \ref{F5}). The latter is smoother, because it contains a larger
number of stars. However, the shape of the annular $N_{\rm A}/N_{\rm
  B}$ profile is consistent with most details in the variations of the
cumulative number ratio profile. For example, if the number ratio in a
given ring is higher (lower) than the average value, the overall value
of the cumulative number ratio at the corresponding radius increases
(decreases) as well.

In summary, our results show that the radial distributions of stars in
the two MS components are not very different. The cumulative profile
of their number and the number ratio distributions do not change
significantly with radius from the cluster's central region to its
outskirts. The annular profile of the number ratio of the blue and red
components, $N_{\rm A}/N_{\rm B}$, is also consistent with the
cumulative profile.

\begin{figure*}
\centering
\includegraphics[width=0.8\textwidth]{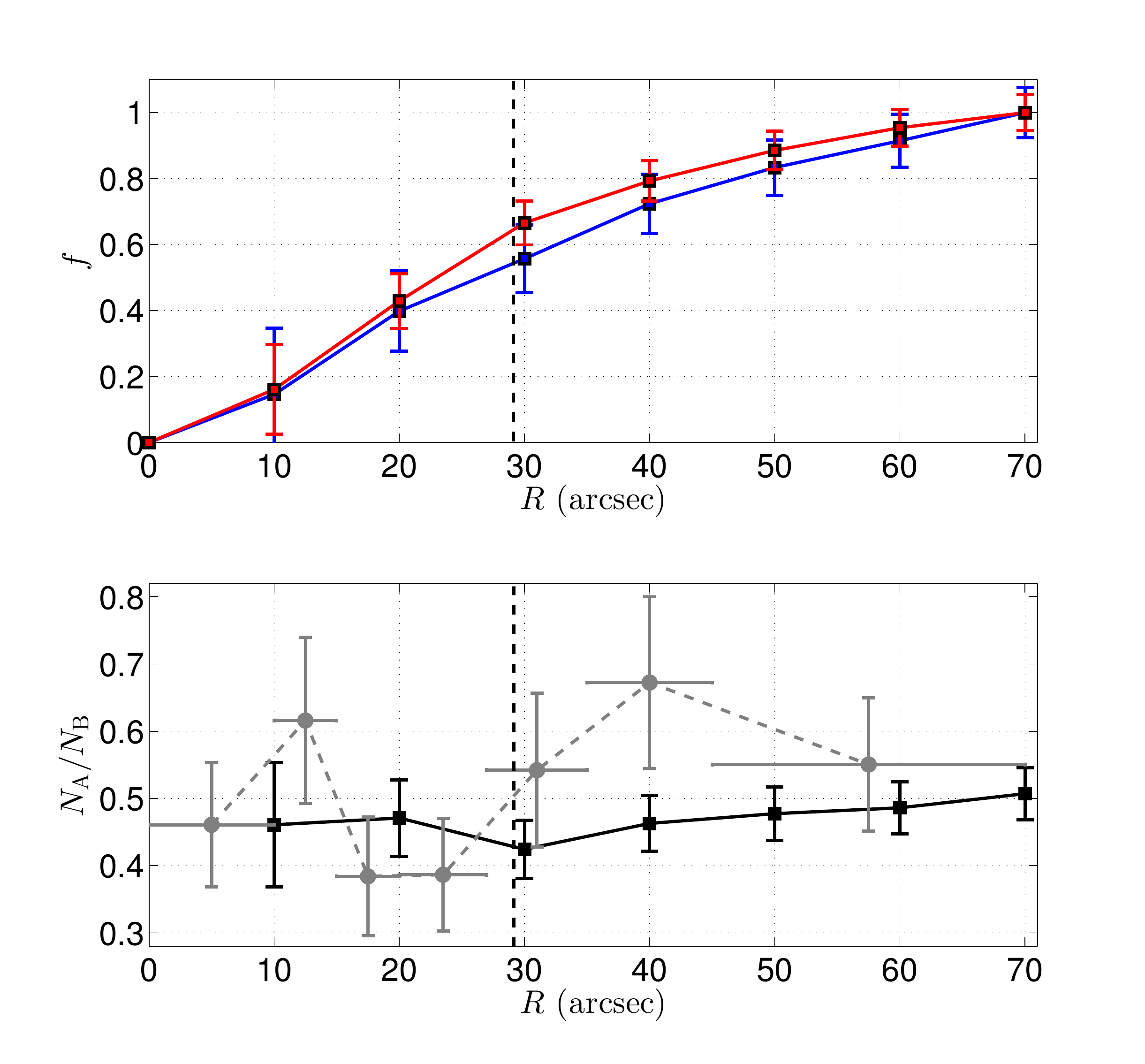}
\caption{(Top) Cumulative number distributions of components A and
  B. (Bottom) Radial profile of the number ratio of components A and B
  ($N_{\rm A}/N_{\rm B}$. Black solid line: cumulative profile; grey
  dashed line: annular profile. The vertical black dashed lines (in
  both panels) indicate the half-light radius of NGC 1856
  \citep{Mcl05a}.}\vspace{5mm}
\end{figure*}\label{F5}

\section{Discussion}\label{S4}
\subsection{Binary Segregation?}

It is unlikely that unresolved binaries are responsible for the
observed MS bifurcation. The color separation between both MS
components in NGC 1856 reaches 0.1 mag or more. Such a large
separation in color indicates that if the red-component stars are
binary systems, they should be approximately equal-mass binaries
\citep[see][their Fig. 9]{Li13a}. However, because the red-component
stars dominate the stellar numbers at a given magnitude, this means
most binaries in NGC 1856 must be high mass-ratio binary
systems. This, in turn, is in conflict with many previous studies
\citep[e.g.,][]{Kou05a,Kou07a,Regg11a}.\footnote{Note that these
  studies are based on explorations of stellar associations.} In
addition, since binaries are more massive on average than single
stars, binaries are usually more dynamically segregated \citep{deg13}
because of long-term two-body interactions \citep[but
  see][]{deg13,Li13a}. If the observed double sequences in NGC 1856's
CMD can be explained by single stars (blue sequence) and binaries (red
sequence), stars that belong to the red MS component should be more
segregated than the blue-population stars. This is in contrast to our
results.\footnote{It is also unlikely that the red-component stars are
  caused by line-of-sight blending. Using artificial star tests, we
  found that the blending fraction pertaining to our sample stars is
  only $\sim$2\%, which cannot significantly change the resulting
  number-ratio profile. In addition, blending should cause the
  red-component stars to be apparently more centrally concentrated
  because of the high number density in the cluster's central region.}

\subsection{Multiple Star Forming Events?}

Two scenarios could potentially explain the MS split in the CMD of NGC
1856: (1) multiple stellar populations of different ages and
metallicities or (2) a coeval stellar population composed of two
stellar components that are distinctly different in terms of their
rotation properties. We first explore the scenario pertaining to the
presence of multiple stellar populations. {\color{black} If more than a
  single star-forming event had occurred, the second-generation stars
  should be characterized by enhancements of both their helium
  abundance and their metallicity. It is well-known that helium
  variations are responsible for split MSs in old Galactic GCs, where
  the multiple sequences are usually well-described by isochrones with
  different helium abundances \citep[e.g.,][]{Piot07a}. For young
  stellar clusters like NGC 1856, the MS regime studied in this paper
  is populated by very hot stars. In this regime of stellar gravity
  and temperature, the MSs of two stellar populations characterized by
  similar metallicities but different helium content are almost
  coincident. This is clearly shown in Fig. 9c of \cite{Milo16a},
  where two isochrones with very different helium abundances ($Y=0.25$
  and $Y=0.40$) are compared with the observed CMD of NGC 1755. This
  figure clearly shows that helium differences cannot reproduce the
  large MS color split observed in NGC 1755, NGC 1856, and other young
  clusters in the Magellanic Clouds.}

{\color{black} A more reasonable assumption is adoption of two stellar
  generations with different ages, helium abundances, and
  metallicities, where the younger-generation stars are both helium
  and metallicity enhanced.}  {\color{black} To test this idea, we
  generated a series of isochrones based on the Padova stellar
  evolution models \citep{Bres12a} for different ages and
  metallicities, with helium abundances defined by $Y=0.2485 + 1.78Z$
  \citep[see][their Section 4]{Bres12a}.} We determined the
best-fitting isochrones to the observations by comparing their
positions with the loci of the adopted ridge lines. We found that
adopting isochrones that are only different in age (but identical in
helium abundance and metallicity) cannot fit the positions of the two MS
components.\footnote{This approach is usually adopted to explain the
  eMSTO regions in most intermediate-age star clusters in the
  Magellanic Clouds \cite[see, e.g.,][]{Milo09a}.} Instead, to
determine the best-fitting isochrones that most closely match the loci
of the adopted ridge lines, we need to adopt isochrones that are
different in both their ages and metallicities. That is, the young
isochrone has a higher metallicity than the old isochrone. Finally, we
determined an age of 290 Myr ($\log (t\ {\rm yr}^{-1}) = 8.46$) and a
(solar) helium abundance and metallicity of $Y=0.275$, $Z=0.0152$ for
the red MS component, while the best-fitting age and metallicity for
the blue-component stars are 340 Myr ($\log (t\ {\rm yr}^{-1}) =
8.54$), $Y=0.263$ and $Z=0.008$ ($\sim 0.5 Z_{\odot}$), respectively.
The extinction and distance modulus adopted for both isochrones are
$E(B-V)=0.15$ mag and $(m-M)_0 =18.35$ mag \citep[cf.][their
  Fig. 9]{Milo15a}. The fits resulting in isochrones with different
ages and metallicities are shown in the left-hand panel of
Fig. \ref{F6}.

\begin{figure*}
\centering
\includegraphics[width=1.0\textwidth]{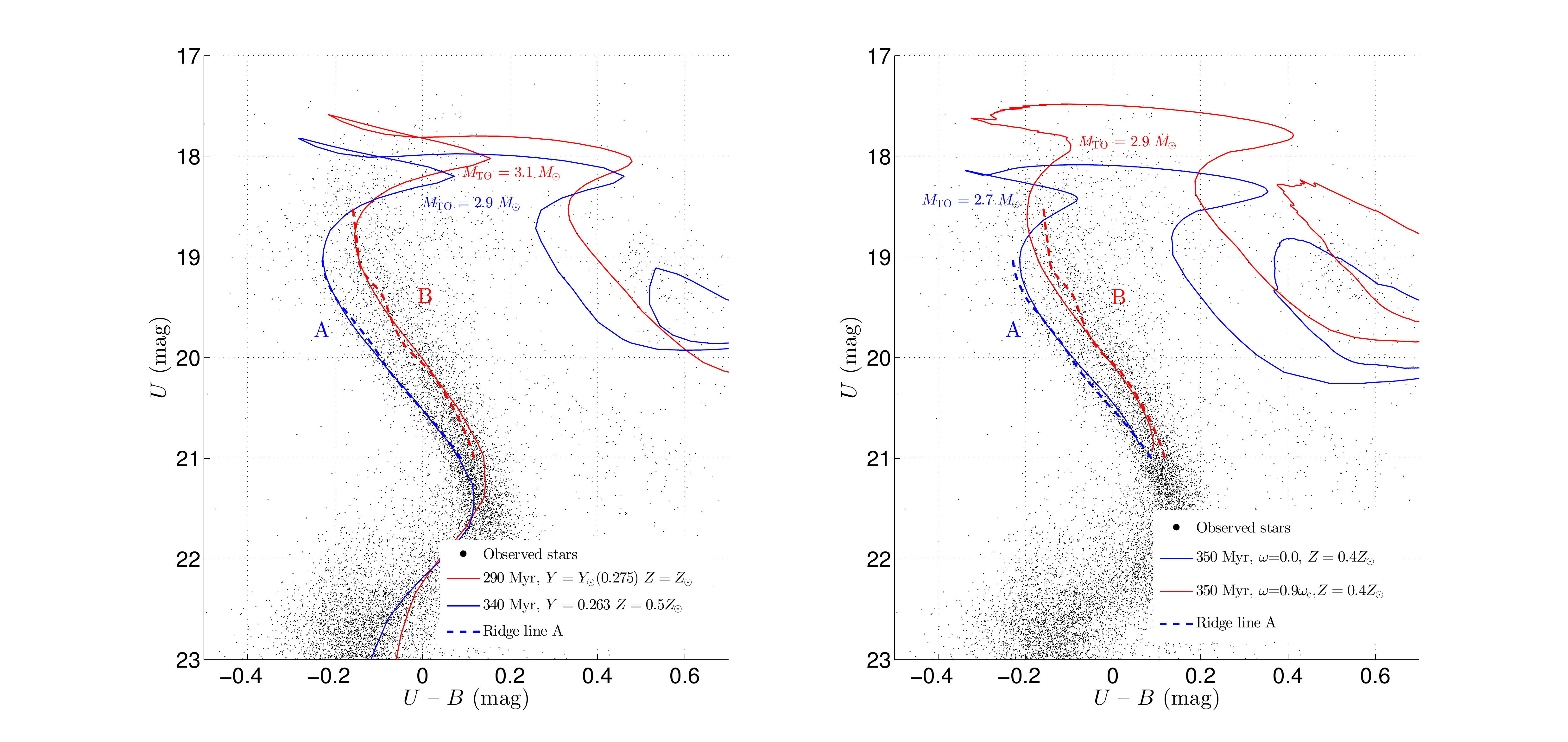}
\caption{Isochrone fits to the two different MS components in NGC
  1856. (left) \cite{Bres12a} isochrones with different ages, helium
  abundances and metallicities, as indicated. (right) Both isochrones
  have the same age and metallicity, but they are characterized by
  different rotation rates (blue, red solid lines: non-rotating,
  rapidly rotating stars with $\omega=0.9\omega_{\rm c}$. These
  isochrones were calculated using the Geneva code
  \citep{Ekst12a}. The blue and red dashed lines are the ridge lines
  for both MS sequences. {\color{black} We have included the stellar
    masses of the turn-off (TO) stars in both panels. All stars
    studied in this paper are B- and early F-type stars.}}\vspace{5mm}
\end{figure*}\label{F6}

If our fits are indeed reliable, then these results indicate that the
red stellar population is actually the second stellar generation,
which formed more recently. This may further indicate that NGC 1856
could have retained the initial runaway gas from supernova explosions,
since the two observed stellar populations are different in their
[Fe/H] abundances. The typical velocity of the runaway gas driven by
Type II supernovae is on the order of 100 km s$^{-1}$ \citep{Roge13a}. 
If we assume that the half-mass radius of NGC 1856
is roughly equal to its half-light radius, $r_{\rm hl}=r_{\rm h}=7.76$
pc \citep{Mcl05a}, and that this also represents the initial
half-light radius of NGC 1856, then the minimum mass for NGC1856 to
retain the escaping gas driven by Type II supernovae is
\citep{Geor09a}
\begin{equation}
M_{\rm cl}\approx100v_{\rm esc}^2r_{\rm h}.
\end{equation}
Here we simply define $v_{\rm esc}\sim$100 km s$^{-1}$. Finally, we
conclude that NGC 1856's initial mass should be {\it at least} $M_{\rm
  cl}\sim 7.76 \times 10^6 M_{\odot}$ to retain the runaway gas driven
by Type II supernovae. The current mass of NGC 1856 is of order $10^5
M_{\odot}$ \citep{Mcl05a}, while the enhanced metallicity of the
second stellar generation in NGC 1856 would indicate that it has lost
more than 99\% of its initial mass during the last 340 Myr. It is
difficult to explain how a cluster can lose such a large amount of
mass during such a short period. This is equivalent to the so-called
`mass budget' problem that affects most scenarios that aim to explain
the observed multiple stellar populations in GCs \citep{Bast15b}
\footnote{Note that \cite{Lar12a} has estimated that the initial mass 
of GCs in the Fornax dwarf spheroidal galaxy would only need to be 
4 -- 5 times more massive than their current mass, on average.}.

In addition, the observed homogeneity in the two stellar populations'
spatial distributions is also different from that seen in most
observed GCs, where the second stellar generation stars usually have a
greater central concentration \citep{Lard11a,Milo12a,Mass16a}. Our
results would be in conflict with most scenarios that invoke the
ejecta of the first-generation stars as the origin of any secondary
stellar generations \citep{Renz08a,Vesp13a,Hnau15a,Khal15a}. All these
scenarios predict that the secondary stellar generation stars should
preferentially form in cluster's central region and thus have a higher
concentration than the first-generation stars. However, a recent paper
has shown revised spatial distributions for some subpopulations in the
GCs NGC 362 and NGC 6723 \citep{Lim16a}, indicating that a higher
concentration of second stellar generation stars may not necessarily
be a common feature of GCs.

If the proposed multiple stellar population scenarios were on the
right track, one possible explanation to the homogeneity in the
spatial distributions of these two sequences is that they may
initially have been different in terms of their radial distributions
but were subsequently mixed through two-body relaxation. However, the
half-mass relaxation timescale for stars in NGC 1856 is $t_{\rm hm}
=2.2$ Gyr \citep{Mcl05a}, which is roughly seven times longer than the
isochronal age of the cluster. This means that the observed spatial
distributions of the two stellar populations should be close to their
initial distributions, that is, they formed with uniform spatial
distributions. However, all these speculations are based on the
assumption that NGC 1856 is actually a young counterpart of the old
GCs. It is also possible that the results from NGC 1856 cannot provide
any constraints to the scenarios for multiple stellar populations in
old GCs, given that all scenarios we have discussed deal with old GCs
\citep{Renz08a,Vesp13a,Hnau15a,Khal15a}.

Based on all the considerations, we conclude that the MS bifurcation
in NGC 1856 is most likely not caused by multiple stellar populations
of different ages and metallicities. 

\subsection{Rapid Stellar Rotation?}

\cite{Dant15a} explored the impact of rapid stellar rotation to
explain the observed split in the MS of NGC 1856. They found that the
observed CMD of NGC 1856 can be interpreted as a coeval stellar
population composed of one-third of slowly/non-rotating stars and
two-thirds of very rapidly rotating stars. We found that their
conclusion is also valid for our (F336W, F336W--F438W) CMD: using the
same isochrones as those adopted by \cite{Dant15a}, we show the
performance of our isochrone fits in the right-hand panel of
Fig. \ref{F6}. We found that the blue and red MS components can be
well described by stellar populations characterized by a zero rotation
rate ($\omega=0$) and a level at 90\% of the critical rotational rate
($\omega=0.9\omega_{\rm c}$), respectively. This means that the number
ratio of slowly/non-rotating and rapidly rotating stars is 39\%/61\%
(see the bottom right-hand panel of Fig. \ref{F5}), which is
consistent with \cite{Dant15a}.

It is unclear whether the slowly/non-rotating stars and their rapidly
rotating counterparts in a star cluster should preferentially form at
different locations. Thus far, no research has explored the
relationship, if any, between stellar rotation rates and their spatial
distributions. In our observations, the rapidly rotating stars are
redder than the non/slowly rotating stars, indicating that the rapidly
rotating stars are strongly affected by gravity darkening
\citep{Li16a,Geor14a}. If gravity darkening is the {\it only} case
that leads to an observed color difference such as that between our
two populations, then this would indicate that the average masses of
the two stellar populations are similar. They would thus expected to
exhibit the same degree of segregation. However, it is unclear at what
level rotational mixing would work for rapidly rotating
stars. Rotational mixing would increase the core size of a rapidly
rotating star, thus extending its MS lifetime compared with that of a
non-rotating star of the same mass. If rotational mixing also plays a
significant role in rapidly rotating stars in NGC 1856, then a less
massive rapidly rotating star would have the same luminosity as a more
massive non-rotating star, because the former has the same core size
as the latter. In this paper, we have constrained our study of the two
stellar components to the same magnitude range, which means that if
rotational mixing is important for the red-component stars, their
average mass should be smaller than that of the blue-component
stars. In turn, this would lead to the former becoming less segregated
than the blue stellar population.

It seems that rotational mixing {\it does} work for NGC 1856. Based on
Figs \ref{F2} and \ref{F6}, we can see these two populations of stars
are not only different in color, but also in the loci of their MSTO
points. The red sequence is characterized by a brighter turn-off
point, indicating that stars belonging to the red MSTO region are more
massive than stars associated with the blue MSTO region. This is
presumably caused by the prolonged MS lifetime owing to rotational
mixing. A promising method to evaluate the importance of rotational
mixing for rapidly rotating stars is to measure the morphology of the
SGBs in star clusters. If rotational mixing is negligible for rapidly
rotating stars, then when stars evolve to the SGB stage, their
rotation rates would quickly decrease because of the conservation of
angular momentum. A coeval stellar population would then display a
tight or converging SGB in its CMD. On the other hand, if rotational
mixing is important, then a rapidly rotating star's core size would be
permanently changed by the mixing effect before it evolved off the MS.
In that case, the rapidly rotating stellar population would appear
brighter than the slowly/non-rotating stellar population on both the
MS and SGB, which would eventually produce a broadened
SGB. {\color{black} Previous studies of intermediate-age clusters, which
  usually exhibit apparent eMSTO regions, have shown that the
  morphologies of their SGBs closely resemble or are consistent with
  simple stellar populations,} indicating that for those clusters
rotational mixing does not dramatically change the core sizes of the
MSTO and SGB stars \citep{Li14a,Bast15b,Li16c,Wu16a}. However, this
still needs to be explored for YMCs.

{\color{black} Finally, one may wonder why the stellar rotational
  distribution exhibits a dichotomy: 66\% of the cluster's stellar
  population appear to be extremely rapid rotators, 34\% are
  apparently non- or slowly rotating stars. \cite{Huan10a} studied
  spectra of more than 230 cluster and 370 field B-type stars. They
  found that most of their least-evolved, low-mass
  ($2M_{\odot}<{M}<4M_{\odot}$) B-type stars are born as rapid
  rotators, while stars with rotation rates $\omega<0.5$ compose 37\%
  of the total population. \cite{Daft13a} studied 334 stars of
  spectral types ranging from O9.5 to B3. They also found a bimodality
  in the distribution of the rotation rates of early B-type
  stars. Both studies are consistent with our results for NGC
  1856. One explanation that may account for the two distinct
  populations in terms of the stellar rotation rates involves binary
  synchronization \citep[see][their Section 4.2]{Dant15a}. In the
  latter case, most slowly or non-rotating stars would be members of
  hard binary systems; the tidal interactions in such binary systems
  are thought to have slowed down the stars.}

In summary, it seems that our results favor the rapid stellar rotation
scenario rather than the multiple stellar populations model. However,
more research of the importance of rotational mixing in rapidly
rotating stars is required.

\section{Conclusions}\label{S5}

In this paper, we have studied the number ratio of stars belonging to
two distinct components of NGC 1856's MS. We found a reliable number
ratio of the blue and red MS components stars in NGC 1856 of
34\%/66\%. In addition, based on the adopted cluster number density
center, both stellar components seem to be distributed homogeneously
in space. Their cumulative number ratio remains constant over the
entire radial range, and only a very weak dispersion appears in a
narrow range for the annular number ratio profile.

We have discussed possible explanations on the basis of unresolved
binary systems, multiple stellar populations, and rapid stellar
rotation. We excluded the unresolved binaries as a cause. We confirmed
that both multiple stellar populations and rapid stellar rotation can
produce the observed features. However, most multiple stellar
population scenarios have problems to explain the homogeneity of the
radial distributions of the two observed stellar sequences. In
addition, the enhanced metallicity that is needed in the multiple
stellar population scenario would inevitably raise the `mass budget
problem' \citep{Bast15b}. We emphasize that this conclusion is based
on the assumption that NGC 1856 is representative of a young
counterpart to the old GCs.

If the dominant effect caused by rapid stellar rotation is gravity
darkening, then the observed MS bifurcation would favor the rapid
stellar rotation scenario. However, currently no comprehensive studies
have explored the relationship, if any, between stellar rotation and
their spatial distributions. In order to shed light on this issue, an
improved understanding of the relative importance of rotational mixing
for rapidly rotating stars is required.


\acknowledgments {We thank M. Di Criscienzo and F. D'Antona for
  providing us with the relevant isochrones. We acknowledge
  L. R. Bedin for his help with the {\sl HST} footprints.  CL is
  supported by the Macquarie Research Fellowship Scheme. CL, RdG and
  LD acknowledge financial support from the National Natural Science
  Foundation of China through grants 11373010, 11633005, and
  U1631102. APM acknowledges support from the Australian Research
  Council through a Discovery Early Career Researcher Award,
  DE150101816.

\bibliographystyle{apj}


\end{document}